\PassOptionsToPackage{table,xcdraw}{xcolor}

\DocumentMetadata{}

\documentclass[sigchi]{acmart}
\usepackage{tabularx}
\usepackage{multirow}
\usepackage{subcaption}
\usepackage{cleveref}
\usepackage{longtable}
\usepackage{array}

\usepackage{hyperref}
\begin{document}

\title {Identifying Ethical Challenges in XR Implementations in the Industrial Domain: A Case of Off-Highway Machinery}

\author{Anastasia Sergeeva}

\affiliation{%
  \institution{University of Luxembourg}
  \country{Luxembourg}}
\email {anastasia.sergeeva@uni.lu}

\author{Claudia Negri-Ribalta}

\affiliation{%
  \institution{University of Luxembourg}
  \country{Luxembourg}}
\email {claudia.negriribalta@uni.lu}

\author{Gabriele Lenzini}

\affiliation{%
  \institution{University of Luxembourg}
  \country{Luxembourg}}
\email {gabriele.lenzini@uni.lu}
\renewcommand{\shortauthors}{Sergeeva  et al.}

\begin{abstract} 
Although extended reality (XR)- using technologies have started to be discussed in the industrial setting, it is becoming important to understand how to implement them ethically and privacy-preservingly. In our paper, we summarise our experience of developing XR implementations for the off-highway machinery domain by pointing to the main challenges we identified during the work. We believe that our findings can be a starting point for further discussion and future research regarding privacy and ethical challenges in industrial applications of XR.

\end{abstract}
%



\maketitle

\section{Introduction}
The integration of extended reality (XR) elements into work contexts is becoming an important component of the transition to Industry 4.0. For example, XR is being implemented in various job processes, such as assembly, maintenance, and industrial training
\cite{wang2024towards,alhakamy2024extended}.
However, while these applications benefit the industry and help improve various aspects of work, they also pose challenges and opportunities that impact the privacy and security of workers and bystanders. 
When discussed within the broader framework of XR, practitioners and scholars often mention the adversarial effects of XR on privacy. This concern arises due to extensive data collection, including sensitive biometric data (e.g. gaze patterns, micromovements of hands) without explicit consent and with purposes that benefit companies but not users.

Implementing XR in the workplace (especially in the context of industrial machinery) introduces additional challenges for developers. Newly developed XR solutions often aim to integrate into existing organizational or technological workflows. This integration requires an additional analysis of existing organizational and technological processes and requirements to ensure that new technologies will not disrupt current operations. Conflicts between the requirements of this new technology and the organizations need to be identified, negotiated, and resolved. 

As XR technologies are still relatively new in industrial applications compared to other sectors, such as entertainment, stakeholders are typically skeptical. This happens especially in industries with previously low levels of digitalization, such as certain off-highway machinery sectors. The low to moderate digitalization levels also imply that introducing XR technologies, such as ``always-on'' sensors or cameras, could significantly increase the collection of personal data from workers and bystanders. This practice introduces changes in the organization's privacy and data protection governance, as the current policy may not cover this new practice.
Moreover, since the collected data becomes intertwined with work processes and can potentially be used to assess employee performance, the extensive data collection enabled by XR creates serious ethical questions about the balance between operational efficiency (business interests) and employees' rights to privacy and other human rights in the working place. 

All the mentioned issues require a systematic approach to identify and specify the main ethics and privacy-related challenges, which can be relevant to specific industries. 
Several research groups and organizations have started to propose and promote general guidelines for the development of XR applications, but only a few have focused on the ethical values and principles that should be integrated into XR-powered solutions \cite{initiative2020xrsi,mcgill2021white}. Moreover,the general XR-related recommendations may not be fully applicable to the specific challenges in the workplace or may have missed some area-specific problems.  Therefore, it is necessary to perform domain-relevant analysis to identify gaps where general guidelines may not suffice for specific challenges. In this manner, collaborative work with industry experts and other stakeholders is vital to identify the gaps.  

In this position paper, we present our view on the main ethics-related challenges for off-highway machinery (e.g. excavators, snow groomers, reach stackers) in the XR domain. Our findings are based on an analysis of the literature on general ethical and privacy guidelines for XR and barriers and challenges of industrial XR. We then conducted pilot studies with end users and experts in off-highway machinery, seeking their opinions about key industry problems. With these preliminary results, we identified a first round of the main challenges of the XR for high-way machinery. We believe that overcoming the issues will help develop ethical and privacy-preserving solutions for the domain and contribute to the general discussion about industrial XR.

The rest of this paper is structured as follows: in \Cref{sec:related-work}, we discuss related work on privacy and XR, which frameworks exist, and XR applications in off-highway machinery. Afterward, in \Cref{sec:challenges}, we present our method used for identifying the challenges (\Cref{subsec:methodology}) and discuss the challenges in \Cref{subsec:challenges-identified}. 

\section{Related Works} \label{sec:related-work}

\subsection{Privacy and the Ethical Values Questions in Extended Reality}

Several studies have shown that XR can pose significant risks to user privacy, security, and their ethical use\cite{giaretta2022security, acquisti2011privacy, paul2014privacy, dokuchaev2020data, khamis2021privacy}. 
The first group of concerns is related to the problem of user identification in XR applications. The applications gather a significant amount of user behavior data --- often without the user's explicit content --- that may lead to the possibility of identifying them \cite{dipakkumar2022systematic, huang2023security}. 
In addition, as many XR applications are tangled to other databases (like profiles on Meta), it is possible not just to identify the user by its behaviour and biometrics, but also link this identified profile to other data types: the name, profile photo or geography \cite{mcgill2021white,huang2023security}. Therefore, it is possible to use this combined profile for multiple purposes \cite{mcgill2021white,huang2023security} while also creating a risk of pushing users to overshare/disclose more personal information, similarly as it happened in non-XR social media domains \cite{giaretta2022security}. 

Overall, as \cite{khamis2021privacy} points out,  unintentional disclosure of confidential data, the complexity of obtaining informed consent, and emerging vulnerabilities for attacks, are critical privacy and security risks in XR. These identified ethical issues can be grouped as surveillance concerns from users to organizations.

In this context, the behavioural information gathered from XR interactions can be used against the user's interests. For example, micromovements data can be analysed from the health perspective, leading to conclusions about the subjects' health, affecting their insurance. In these situations, the XR-based data collection can lead to discriminatory actions towards users \cite{pahi2023extended}. 

On the technological level, most of the studies focus on eye-tracking technology \cite{david2021privacy, partala2000pupillary}, which has been on the market for years and has a clear data-gathering path. The associated risk is that eye-tracking helps collect sensitive user information, such as preferences and mental load, often without explicit consent \cite{partala2000pupillary}. Several studies are dedicated to the problem of privacy in wearables \cite{paul2014privacy,zhang2017electromyogram}. 

On the user side, there are works dedicated to the lack of user awareness regarding data collection, opaque practices by XR developers, and potential manipulative tactics by malicious agents, exploiting these issues and XR's unique characteristics \cite{abraham2022implications,adams2018ethics,halbig2022opportunities,hadan2024privacy}.
Hadan et al. identified factors influencing users' concerns about data privacy in XR \cite{hadan2024privacy}. The system, which collects cognitive, emotional, and personality data, is perceived as ``too intrusive''. The study also indicated an increase in users' discomfort if the XR device operated ``in the background'' as somehow hidden from users' attention focus.  The perception of algorithmic unfairness in data collection also contributed to the level of users' discomfort \cite{hadan2024privacy}.
The study of O'Hagan et al. showed users low awareness but high concern about AR headset activities, particularly those related to bystanders' data \cite{o2023privacy}.

\subsection{Frameworks and Guidelines in Privacy and Ethics of Work in XR}
There are several frameworks or foundational papers which can be considered as frameworks/guidelines in the design of XR solutions with a privacy/ethics focus. The framework and software implementation proposed by Reilly et al. aimed to synchronize privacy-related actions across both physical and virtual spaces \cite{reilly2014secspace}. The main features of their tool includes a shared ontology for physical-virtual interaction, a flexible communication model, and interface mechanisms that support user experience across realities\cite{reilly2014secspace}. 

In another initiative, De Guzman et al. \cite{de2019security} summarised a list of security and privacy concerns and created an ontology, which includes privacy model attributes such as integrity, non-repudiation, availability, authorisation, and authentication. In addition, in their work they divide the areas which should be defended as data input, data itself, data output, user interaction, and device protection.The XRSI Privacy Framework by the XR Safety Initiative focuses on assessing privacy risks, informing users about these risks, managing privacy, and preventing privacy-related incidents in XR settings \cite{initiative2020xrsi}. The XRSI framework aligns its measures with major legal privacy and compliance standards. 

The IEEE Global Initiative on Ethics of Extended Reality\cite{mcgill2021white} identifies key privacy challenges produced by Extended Reality: capturing the sensitive and behavioural data (which leads to the possible penetration of mental privacy), the adversarial effects on bystanders' privacy, total surveillance and effect on the perception of reality and propose a list of recommendations for developers in the form of suggestions from design and legislation perspective\cite{mcgill2021white}. 

Norval et al. discussed how XR systems are linked to their environment and user actions \cite{norval2023navigating}. The physical nature of XR (interactions in the real world) means that problems can lead to immediate and serious consequences, such as injury and property damage, which go beyond the usual data-related issues. Because of the severity of potential outcomes, it is important that the auditors can reconstruct or comprehend the system's behaviour and its context during operation.
Norval et al. developed and tested a framework with experts to assist in the development of the XR system, which aims to facilitate the capture and management of data, a necessary requirement for conducting audits \cite{norval2023navigating}. 
Abraham et al.\cite{abraham2024you} explored the concept of fine-granular permission control for applications in XR. They found that such an approach helps users better understand the balance between system performance and data collection, allowing them to make informed decisions about their privacy \cite{abraham2024you}. 
Finally, the E3XR framework, developed by Lee and Hu-Au \cite{lee2021e3xr}, evaluates XR technologies in three dimensions: ethics, educational efficacy, and eudaimonia (human flourishing). Ethics is defined as user privacy, autonomy, and avoiding harm. Education is described as interactive immersive experiences that enhance learning. Eudaimonia assesses XR's capacity to promote personal growth and social good, as well as inclusivity and empowerment.

\subsection{XR Applications in Industry and in the Domain of Off-highway Machinery}

To date, a limited number of studies and case discussions have been dedicated to the use of XR in the off-highway machinery domain. Even in existing research, XR is mainly focused on a few topics. For example, a systematic review of the literature by Zoleykani et al. found that most studies on XR in construction primarily discuss its application in the design aspects of construction projects, with very little discussion on operational aspects \cite{zoleykani2023extended}. Although there is potential to apply XR in end-user applications such as operator safety training, these applications are considered mainly a subject for future research \cite{zoleykani2023extended}. Similarly, in discussing XR applications in the agricultural domain, which overlaps with off-highway machinery, Fountas et al. mentioned benefits such as the teleoperation of agricultural tractors \cite{fountas2024agriculture}. Nevertheless they also noted that this has not yet been implemented in agricultural processes.

In the broader context of creating XR solutions in manufacturing, previous literature suggests the applicability of certain mixed reality (MR) and virtual reality (VR) approaches in different manufacturing stages. For example, VR can be used primarily in the design phase, while MR technologies can already be applicable in the whole industrial cycle \cite{fast2018testing}. 
In the dedicated literature for augmented reality (AR) implementations, most AR applications in industrial environments focus on providing work instructions for manual assembly processes, maintenance, and training. AR is also used to improve safety in human-robot operations (e.g., by showing the robot's movement trajectory to the human) and in the broader context of on-site analysis, evaluation, and decision-making. \cite{de2020survey}.
Vasarainen et al. showed a growing discussion on using XR to improve existing working practices, especially in stressful and dangerous situations. 

However, certain existing limitations, such as the level of precision and seamlessness of interaction, make the use of XR problematic in some professional fields. Furthermore, the relative novelty of the technology can cause operators to focus more on the virtual aspect of the interaction, potentially neglecting the real-world situation \cite{vasarainen2021systematic}. 

Summarizing the discussed findings, we can conclude that while both industry and academia agree on the importance of XR technologies for the further development of industrial processes, including those related to off-highway machinery, the area is still underdeveloped and requires both further technological improvements and extensive user studies to understand the conditions of XR technology acceptance in the industrial context.

\section{Identifying the main privacy and ethical challenges for XR for Off-Highway Machinery} \label{sec:challenges}

\subsection{Methodology} \label{subsec:methodology}
To take the fist steps in the identification of the challenges for ethical XR implementation in the off-highway machinery domain, we applied the following procedure:
\begin{itemize}
    \item We conducted a series of workshops in the frames of Value-Sensitive Design \cite{friedman1996value}, the list of values we used was based on a short version of Schwartz values in the working context, built on   \cite{schwartz2012overview, lindeman2005measuring, albrecht2020measuring}  with end-users (n = 11) to determine their perceived needs for privacy and ethical work; we also organized three workshops with other stakeholders and developers in the project to identify relevant privacy challenges in the development of XR solutions using the Linddun Go methodology \cite{wuyts2020linddun} and GDPR-based questionnaire (n = 10); summary of our findings is presented in the Appendix;
    
    \item We conducted additional in-depth interviews with industry experts (n = 3) in each of our use case domains to better understand the state-of-the-art privacy and ethics in the industries and align future technological visions with the need to preserve privacy and ethics. A brief summary of answers is added to the Appendix; 
    
    \item We triangulated the results of the expert interviews, stakeholders' privacy discussions, and analysis of privacy-related frameworks and challenges in XR, mentioned in the Related Work section.
\end{itemize}

\subsection{Relevant Privacy and Ethics Challenges of XR Development in the Off-Highway Machinery Context} \label{subsec:challenges-identified}

Based on our work, we identified key challenges that we found most relevant to XR for Off-Highway Machinery development. This is a preliminary list of identified challenges that aims to serve as a starting point but does not intend to be exhaustive nor cover all XR challenges. 

\subsubsection{Presence of Bystanders at the Operation Sites}
As mentioned in \cite{mcgill2021white,pahi2023extended}, it is critical to protect the privacy and identity of individuals in locations where XR tracking is operational. Although the captured data may not inherently violate privacy expectations, previous work shows that when combined with existing extensive databases (such as social media platforms\cite{huang2023security}), these data pose significant privacy risks. In the case of off-highway machinery, the risk of collecting bystander data is high, as XR-equipped machines are planned to operate in semi-enclosed areas of everyday city landscapes where other workers or the general public may be present and captured by XR solutions without their explicit consent. In these cases, bystanders are probably not aware of the presence of the technology, as they have just started to be introduced in industrial settings, so their understanding of privacy risks is expected to be very limited. 
While previous works discussed some ways to inform the bystanders about the data collection in the area \cite{mcgill2021white}, the industry still not yet discussing their applicability in the context. 

An additional challenge in this respect can be that on the construction site, it can be difficult to draw a clear line between bystanders, who have legitimate rights to be on the place and trespassers, who are violating the safety (and possibly legal) requirements by being here. The situation poses a significant ethical question: should the technologies, by default, protect the privacy of trespassers, while the collected information should help prevent the operation site requirements violations?
\\

    \fbox{\parbox{7.8cm}{\textbf{Challenges identified:}
    \begin{itemize}
        \item How can a bystander be informed of the usage and tracking of XR technologies? 
        \item How to differentiate between bystanders and trespassers?
        \item What are the regulatory challenges for privacy and data protection in this context?
    \end{itemize}}}

\subsubsection{State of the Art for Digitalization in the Industry and Existing Business Processes}
The interviews with experts showed that the level of digitalization in different areas of off-highway machinery is still not high and may distinctly vary from one service provider to another. In all three discussed use cases, the experts acknowledged the ongoing discussion about common data-sharing protocols and a standardized approach to developing technologies, which is still a continuous process. Despite the interest, existing and widely shared XR implementations exist in the industry.

This lack of standardisation implies that during the digitalization process of off-highway machinery, including XR enhancement, it is important to avoid disrupting the business processes and ensure their continuity and operability of them. 
This situation poses the question of how to seamlessly integrate XR processes into existing business processes within an organization. From the point of view of privacy and ethics, it can be formulated as follows: a) how can we ensure that added XR practices do not create additional privacy risks compared to the existing baseline of the working process\footnote{At this point, this process mostly does not collect any personal data due to the low digitalization of vehicles.} and b) how can we be sure that XR implementation will not create risks of unethical data use (e.g., performance evaluation without the worker's knowledge and consent) or unethical workflow (e.g., over-guidance or provocation of overreliance). The question was pointed out by our end-users as one of their concerns, which is also in line with previous expert study of  Ursin et al. \cite{ursin2024intraoperative} (in the context of a medical setting, which requires similar levels of concern).
\\

    \fbox{\parbox{7.8cm}{\textbf{Challenges identified:}
    \begin{itemize}
        \item How do we take into account the different levels of digitalization in the domain?
        \item How to integrate XR technologies, without disrupting the business and creating new privacy and ethical risks?
    \end{itemize}}}

\subsubsection{24-Hours a Day Data Collection}
The multiple sensors and cameras in XR setups can produce increased privacy risks \cite{mcgill2021white}, which are relevant in the case of off-highway machinery sensors as they usually running during full operation. Additionally, areas like container terminals or construction sites are usually monitored 24 hours a day to ensure the security of the space. This usage can generate concerns of mass surveillance and data gathering, which, in the augmented reality case, raises problems with capturing private data from different actors.
\\

    \fbox{\parbox{7.8cm}{\textbf{Challenges identified:}
    \begin{itemize}
        \item Can security requirements align with data minimization goals? 
        \item How can a bystander apply their data subjects rights over personal data?
    \end{itemize}}}

\subsubsection{High Cost of Overreliance and Other Types of Cognitive Mistakes While Operating the Vehicle}
When improperly operated, off-highway machinery can pose serious risks to the general safety of drivers and bystanders. The issue of misinterpretation of XR information in this setting becomes extremely dangerous. For example, previous studies have shown the possibilities of cybersecurity attacks, which can trigger user mistakes and lead to potential physical harm \cite{casey2019immersive}. 

However, misguidance is also possible in the case of system errors without a malicious agent behind it. During discussions with end-users about possible implementations, experienced operators expressed concerns about situations where students extensively rely on features of XR systems without considering other parameters (e.g., machine indicators and weather conditions). Therefore, they might not be able to identify situations where systems misbehave or provide false information. This could lead operators into potentially dangerous situations as
they might over-rely on the XR system's outputs without critically assessing the accuracy or relevance of the information. The lack of cross-referencing with real-world indicators could result in hazardous decision-making, leading to harm to the operator and damaging the vehicle.
\\

    \fbox{\parbox{7.8cm}{Challenges identified:
    \begin{itemize}
        \item What practices and design choices should be promoted to avoid over-reliance and encourage critical thinking?
        \item Which design patterns (in the context of GUIs) should be promoted to support appropriate levels of trust? 
    \end{itemize}}}

\subsubsection{Use of AI Models to Improve the Algorithms of XR Solutions Operation} 
There are multiple ways AI can be used in the domain of XR solutions for off-highway machinery. For example, it can be used in developing and updating algorithms for collision avoidance \cite{aizat2023comprehensive,muzahid2023multiple}, optimizing user control and feedback during teleoperation, and providing real-time suggestions for performance optimization by guiding the use of visual and XR cues \cite{luo2024user}. Significant progress has been made in the domain of industrial robot teleoperation \cite{marino2020ai} and the operation of autonomous vehicles\cite{ma2020artificial}. 
Yet, in the case of XR  technologies for off-highway machinery, it is necessary to at least fine-tune the existing algorithms based on the data from real operations. As it can require recording an excessive amount of data from the operation side, this can lead to storing additional data after the operation. These data can include bystander data, which were collected without their explicit consent. Moreover, it would not be possible to take the data out of the system if the bystanders requested it. 
\\

    \fbox{\parbox{7.8cm}{\textbf{Challenges identified:}
    \begin{itemize}
        \item How to fine-tune the model, while respecting bystander privacy?
        \item What are the regulatory challenges for AI, with respect to data protection regulations?
    \end{itemize}}}

\section{Conclusion}
While XR technologies are becoming more and more in use in different industrial domains, their use in the off-highway machinery domain is still not omnipresent. This creates a challenge in developing them within the frameworks of privacy-preserving and ethical work practices. Although there are well-developed initiatives covering various ethical and privacy aspects in XR, we believe that in the future more specific guidelines addressing the unique aspects of XR applications in the off-highway machinery domain can help developers and stakeholders. To develop these guidelines, it is necessary to focus on the main industry-specific challenges. We hope that the challenges we identify will contribute to future work in developing guidelines for off-highway machinery.

\begin{acks}
This paper has received funding from the European Union Horizon 2020 research and innovation programme HORIZON-CL4-2022-HUMAN-01-14 (TheiaXR), the Luxembourg National Research Fund REMEDIS (REgulatory and other solutions to MitigatE online DISinformation - INTER/FNRS/21/16554939) and from the European Uninion Horizon 2020 research and innovation programme under the Marie Skłodowska-Curie Actions grant agreement No. 101081455 (OBI-PIA) link \url{https://yia.uni.lu/}.

\end{acks}
\bibliographystyle{ACM-Reference-Format}
\bibliography{REF}

\appendix
\section{Brief summary of stakeholders and end-users considerations about privacy and ethics of work}
\subsection{Data Protection Considerations}

In all three use cases, the stakeholders have privacy concerns around the:

\begin{itemize}
    \item \textbf{Scoring and Evaluation Based on Data}: Stakeholders agreed that the current workflow operator's performance could be already assessed and used for making administrative decisions; however, they are afraid that in  XR-enhanced workflow, the data could be used for in-depth performance analysis.
    \item \textbf{Systematic Monitoring}: Planned features, such as 360-degree cameras, drone cameras, and thermal cameras at workplaces and on machinery,can be used as tools for non-stop monitoring of the workspace and operators' actions. The surveillance on the operation site (e.g. construction site) can also be viewed as systematic monitoring of public areas.
    \item \textbf{Large-Scale Data Processing}: The XR-enhanced solutions can generate a lot of data about vehicle behaviour and operator response via information collected by head-mounted and haptic devices. Stakeholders agreed that it is important to determine which data should be retained and which should be deleted during operations so as not to create a large amount of sensitive data stored.
    \item \textbf{Matching Datasets}: As data collection in the XR-enhanced workflow comes from various sources (e.g. it can be matched sets of information coming from mounted 360-degree cameras, thermal cameras and collision avoidance systems), it can give more opportunities to identify a person (other workers or bystander/trespasser). Stakeholders agreed that the project should implement measures to prevent data matching.
\end{itemize}

\subsection{Technologies-specific risks}

\begin{itemize}
    \item \textbf{Object Detection Algorithm's Training}: The stakeholders proposed to use thermal cameras to identify trespassers/bystanders; in this case, they are planning to implement AI algorithms for better object recognition and prediction of the object movements. This led to the data collection of the trespassers/bystanders without their explicit knowledge and consent. Moreover, as the data will be used for further algorithm training, it will complicate the situation by completely erasing the data if the trespasser/bystander would like to do so.
    \item \textbf{Teleoperations-Supporting Technologies}:Teleoperation-supporting technologies create risks of data transmission from the vehicle to the operation spot. Right now, the data are stored in the vehicle; however, in the teleoperation case, they'll be stored somewhere else, probably in the cloud. Also, the implementation of teleoperation could require more data to be collected (e.g., more data about user feedback).
    \item \textbf{Technologies, Capturing Performance Data}: The stakeholders see the tension between the technologies which allow behavioural data collection and users' privacy. From one side, most of the proposed implementation can provide data about user performance and behaviour, which can help the user to recognise their errors and generally help to learn from their experience. On the other hand, the collection of data can be the path to privacy infringement, operators' concerns about assessment, and operators' stress.
\end{itemize}

\subsection{End-Users Considerations}
Most users reported limited interest in reusing and storing the data of the operation process; they also did not see specific privacy and security risks in the operations. However, some of them expressed concerns about promotion-related assessments based on performance data. Additionally, they mention that sharing the performance information with colleagues would be problematic because it can create unnecessary competition between workers (and provide additional issues with security in the workplace).
 
The operators most often mentioned self-direction, security and team connections (via benevolence/conformity values); they specifically stressed the importance of independence in both decision-making in operations and in the context of being able to make the operation decisions without over-relying on technology. Several operators pointed to the fact that, in some cases, new XR-provided information can become a substitute for "real" information, and that can lead to operational errors. Also, the operators mentioned that if the XR feature can have errors and start to provide false information, in that case, the operator should be able to verify the information by using other measures; however, they are concerned that junior operators may not be able to identify the exact moment they should start verifying the information.

\section{Summary of Experts Interviews About the State-of-Art in Ethics and Privacy in the Different Domains of Off-Highway Machinery}
\onecolumn
\begin{longtable*}{|p{5cm}|p{3cm}|p{3cm}|p{3cm}|}
    \hline
    \textbf{Question} & \textbf{Snowgroomers} & \textbf{Reachstackers} & \textbf{Excavators} \\ \hline
    \endfirsthead
    \hline
    \textbf{Question} & \textbf{Snowgroomers} & \textbf{Reachstackers} & \textbf{Excavators} \\ \hline
    \endhead
    \hline
    \endfoot
    \hline
    \endlastfoot
    Does the vehicle collect (or can potentially collect, e.g., the feature exists in the new vehicles but was not present in the older models) some types of information such as: & ~ & ~ & ~ \\ \hline
    GPS data & yes, it is mapping the position with GNSS (Global Navigation Satellite System), like GPS, Glonass, Galileo or Baidu & yes & some of the vehicles can map position with GPS, but not the old vehicles \\ \hline
    data about the speed of movement and directions of the movement & yes & yes & yes \\ \hline
    environmental conditions (like temperature, moisture) & weather condition information coming from station, there is telematic system for temperature update which store no data & no & no \\ \hline
    (any) obstacle and terrain information/terrain analysis through sensors or cameras (e.g. position of other objects nearby), borders of operation site & no for terrain analysis, some – for obstacle detection & no & partly, some excavators have cameras \\ \hline
    operational efficiency metrics (e.g. fuel usage) & yes & yes, the processes connected to the machine performance & not any of them, but starts to become more popular, helping determine the real use of the vehicle (and prevent personal use situations) \\ \hline
    safety-related data (proximity alerts and collision warnings) & In development & the system is not collecting them, just inform operator; but collect collision information & no \\ \hline
    If any of that is collected, where and how this information is typically stored? How long is it usually stored? If there any additional rules for longer storage in case of incidents? & It is stored in the cloud; operating companies make agreements with the snow groomer developers; there are no specific rules for the information about incidents. & no special rules for incident handling, the collision avoidance call given, but not specifically stored; information is stored in the cloud for big companies, smaller companies save the information locally. & for now, most operational data are stored in the excavators themselves. It is possible to load this data onto the cloud, but using Wi-Fi can cause network connection issues \\ \hline
    How does authorization for vehicle use occur, step-by-step? Where is the authorization data stored? & anonymised data stored on the vehicle developing company in pseudonymised form (the pseudonymisation should be performed on the operating company side); the keys are stored on operating companies’ level & the operating company assigns the individual operator a number. The machine development company only has the operator's number and does not know the identity of the operator. & in many companies there is no specific association between user and vehicle (any operator can use any key to authorise, the area is not very digitalised), however more modern vehicles are identified through login procedures \\ \hline
    How is access to the data typically managed? Have the companies considered scenarios where operators request access to their data (if you ever heard about these types of requests)? If so, what types of data might they inquire about? & no information about such requests, data are anonymous for vehicle developing company, theoretically operating company can request delete/provide data. & no information about such requests; companies providing data to terminals (KPIs about the customer operations, fuel level or tyre pressure) & no information about such requests \\ \hline
    Does the vehicle monitor any vital signals from the operator (e.g. fatigue levels/ alertness through eye tracking)? If yes, where does the information store? & no & no & no \\ \hline
    If the operator leaves the company, what usually happens to the data associated with them (e.g., performance data, if they are collected)? & vehicle data are stored; in case of mapping to operator, the standard setting is stored in anonymous version. & if any data stored, they are either anonymised (no connection to operator) or pseudonymise (company can relate them to the operator), but by default they are staying in the company. & if any data stored, they are either anonymised (no connection to operator) or pseudonymise; they are staying in the company. \\ \hline
    Have the companies usually implemented any analytical approaches to the data? Do the data used for promotion/salary decision? Do the data use for other purposes, e.g. AI models training? & yes, but not any personal operator data & yes, the companies can use anonymized data for analysis and training AI; from the operator efficiency metrics side there is not yet requests from the companies about this type of service & use of AI is just starting in the industry, but it is probably the way forward for bigger companies, as it can help with optimization. However, it will not be possible to use this data for promotion decisions, as the trade unions will probably not approve it. \\ \hline
    If there are any standard consent forms in the industry, explaining the operators GDPR related issues connected to their work, or each company implement their own forms? & no information, happened on the operating company side & for a moment there are initiatives in industry about using common protocols for data sharing, but it is rather on terminal side and it relates to machine data & decided on business level, e.g. there is not yet fully applied data sharing protocols, and all the discussion is related to machine data. \\ \hline
    Are there any fleet management systems used in vehicles? If yes, how do they work, what types of data are collected, and how is the data stored (e.g., is the cloud hosting internal to the company or external)? Can end-users (operators) choose whether to use fleet-management-related features, or is their use mandatory in the contract? & at least some companies implement fleet management features, information stored in encrypted cloud, there is no opt-out options. & It can be useful to coordinate the machine and operator; by default, the collected information cannot be switched off & it is technically possible and there are several solutions on the market, which aim to coordinate the vehicles; in a moment the system is more to assign vehicle to task and not for coordinating operators. They can call each other if they need coordination \\ \hline
    What is the current industrial opinion about teleoperation/fully automated operations? Is it already much in use? What is the horizon of planning for intense automation in the field? What are the main barriers (e.g., technical, social etc.) for that? & still the question for the future, the surrounding is too unpredictable. & there are teleoperations in the container handling industry for larger containers (automated terminals) operating machines, but not yet for reach stackers. The surrounding environment and the type of operations performed in the terminal are more difficult to automate compared to autonomous driving. & still the question for the future, the surrounding is rather unpredictable (compare to the road, where task for cars is easier to implement). \\ \hline
\end{longtable*}

\end{document}